\documentclass[twocolumn]{aastex631}

\usepackage{newtxtext,newtxmath}
\usepackage[T1]{fontenc}
\usepackage{ae,aecompl}

\usepackage{graphicx}    % Including figure files
\usepackage{amsmath}    % Advanced maths commands
\usepackage{enumerate}
\usepackage{hyperref}

\newcommand{\aref}[1]{\hyperref[#1]{Appendix~\ref{#1}}}

\hypersetup{linkcolor=red}          % red equations
\shorttitle{High-$z$ discs?}
\shortauthors{Wang et al.}
\graphicspath{{./}{figures/}}
\begin{document}

\title{On the kinematic nature of apparent discs at high redshifts:

Local counterparts are not dominated by ordered rotation but by tangentially anisotropic random motion}

\author[0000-0002-6137-6007]{Bitao Wang}
\affiliation{School of Physics and Electronics, Hunan University, Changsha 410082, China}
\affiliation{Kavli Institute for Astronomy and Astrophysics, Peking University, Beijing 100871, China}

\author{Yingjie Peng}
\affiliation{Kavli Institute for Astronomy and Astrophysics, Peking University, Beijing 100871, China}
\affiliation{Department of Astronomy, School of Physics, Peking University, Beijing 100871, China}

\author{Michele Cappellari}
\affiliation{Sub-department of Astrophysics, Department of Physics, University of Oxford, Denys Wilkinson Building, Keble Road, Oxford OX1 3RH, UK}

\author[0000-0003-1015-5367]{Hua Gao}
\affiliation{Institute for Astronomy, University of Hawaii, 2680 Woodlawn Drive, Honolulu HI 96822, USA}

\author{Houjun Mo}
\affiliation{Department of Astronomy, University of Massachusetts, Amherst MA01003, USA}
\affiliation{Tsung-Dao Lee Institute, Shanghai Jiao Tong University, Shanghai 200240, China}

%% Note that the \and command from previous versions of AASTeX is now
%% depreciated in this version as it is no longer necessary. AASTeX
%% automatically takes care of all commas and "and"s between authors names.

%% AASTeX 6.31 has the new \collaboration and \nocollaboration commands to
%% provide the collaboration status of a group of authors. These commands
%% can be used either before or after the list of corresponding authors. The
%% argument for \collaboration is the collaboration identifier. Authors are
%% encouraged to surround collaboration identifiers with ()s. The
%% \nocollaboration command takes no argument and exists to indicate that
%% the nearby authors are not part of surrounding collaborations.

%% Mark off the abstract in the ``abstract'' environment.
\begin{abstract}

It is not straightforward to physically interpret the apparent morphology of galaxies.
Recent observations by James Webb Space Telescope (JWST) revealed a dominant galaxy population at high redshifts ($z>2$) that were visually classified as discs for their flattened shapes and/or exponential light profiles.
The extensively accepted interpretation is that they are dynamically cold discs supported by bulk rotation.
However, it is long known that flattened shapes and exponential profiles are not exclusive for rotating disc structure.
To break degeneracy and assess the rotational support of typical high-$z$ galaxies in the JWST samples, those with active star formation and stellar masses $\mathrm{lg}(\mathcal{M}_{\star}/\mathcal{M}_{\odot})\sim9$, we study the kinematics of their equal-mass counterparts at $z=0$.
While these local star-forming low-mass galaxies are photometrically similar to real dynamically cold discs, they are not supported by ordered rotation but primarily by random motion, and their flattened shapes result largely from tangential orbital anisotropy.
Given the empirical and theoretical evidence that young galaxies are dynamically hotter at higher redshifts, our results suggest that the high-$z$ JWST galaxies may not be cold discs but are dynamically warm/hot galaxies with flattened shapes driven by anisotropy.
While both having low rotational support, local low-mass galaxies possess oblate shapes, contrasting the prolate shapes (i.e. cigar-like) of low-mass systems at high redshifts.
Such shape transition (prolate$\Rightarrow$oblate) indicates an associated change in orbital anisotropy (radial$\Rightarrow$tangential), with roots likely in the assembly of their host dark matter halos.

\end{abstract}

%% Keywords should appear after the \end{abstract} command.
%% The AAS Journals now uses Unified Astronomy Thesaurus concepts:
%% https://astrothesaurus.org
%% You will be asked to selected these concepts during the submission process
%% but this old "keyword" functionality is maintained in case authors want
%% to include these concepts in their preprints.

%% From the front matter, we move on to the body of the paper.
%% Sections are demarcated by \section and \subsection, respectively.
%% Observe the use of the LaTeX \label
%% command after the \subsection to give a symbolic KEY to the
%% subsection for cross-referencing in a \ref command.
%% You can use LaTeX's \ref and \label commands to keep track of
%% cross-references to sections, equations, tables, and figures.
%% That way, if you change the order of any elements, LaTeX will
%% automatically renumber them.
%%
%% We recommend that authors also use the natbib \citep
%% and \citet commands to identify citations.  The citations are
%% tied to the reference list via symbolic KEYs. The KEY corresponds
%% to the KEY in the \bibitem in the reference list below.

%%%%%%%%%%%%%%%%% BODY OF PAPER %%%%%%%%%%%%%%%%%%

\section{Introduction}\label{sec:intro}
% previous achievements at lower z.
% looking forward at high z.
% the expectation from simulations.
% JWST finds the disc dominance at high z, assuming flattened objects with exponential profiles are oblate discs.
% flattening of giant ellipticals due to anisotropy.
% the link between flattening exponential profiles and discs.
% this works shows the particular case of low-mass galaxies
% combining other evidence from observations and simulations we speculate the real case for galaxies at high z.

Over the course of understanding galaxy populations in the Universe, it has been a long-lasting pursuit to measure the intrinsic three-dimensional shapes and the internal kinematics of galaxies.

It was not widely realized until the 1920s that galaxies are stellar systems beyond the confine of our Milky Way.
Just a decade after, Hubble's morphological classification scheme was introduced \citep{1936rene.book.....H} and has been serving as the basis even for modern studies related to galaxy morphology.
Hubble's classifications acknowledge the diversity in galaxy intrinsic shapes, in the sense that early-type elliptical galaxies are more spheroidal in three dimensions (i.e. large minor-to-major axis ratios) and late-type spiral galaxies are oblate discs \citep{1970ApJ...160..831S}.
The existence of intrinsically spheroidal galaxies is supported by the fact that the distribution of projected galaxy ellipticity in the local Universe contains too many round objects to be explained by randomly oriented discs alone \citep{1926ApJ....64..321H}.

With statistical analyses of projected shapes of galaxies and their subcomponents \citep[e.g.,][]{2008MNRAS.388.1321P, 2009ApJ...706L.120V, 2010MNRAS.406L..65S, 2013MNRAS.434.2153R, 2020ApJS..247...20G, 2024arXiv240719152X} and of stellar kinematics \citep[e.g.,][]{2011MNRAS.414..888E, 2011MNRAS.414.2923K, 2013MNRAS.432.1709C}, now it is generally believed that local galaxy populations dominant in mass are mainly axisymmetric oblate and spheroidal systems, and that the most massive early-type galaxies are mildly triaxial \citep{2014MNRAS.444.3340W, 2018ApJ...863L..19L}.
The fast-rotating oblate and slow-rotating spheroidal/ellipsoidal galaxies manifest a bimodal distribution of their rotational support \citep{2016ARA&A..54..597C, 2018MNRAS.477.4711G, 2021MNRAS.505.3078V}, which has recently been shown to be a universal property for galaxies across different masses, star formation states, and environments \citep{2023ApJ...950L..22W, 2023NAL000W}.

At redshift $z\sim0$, dynamically hot spheroidal/ellipsoidal galaxies only make up the minority unless galaxies are overly massive ($\mathcal{M}_{\star}\gtrsim10^{11}\mathcal{M}_{\odot}$) and with relatively low star formation rates \citep{2020MNRAS.495.1958W, 2021MNRAS.503.4992F}.
Modern simulations predict that hot spheroids are much more prevalent at higher redshifts before the cosmic noon $z>2$, when even the star-forming galaxies are generally far from being dynamically cold \citep[EAGLE, TNG100, TNG50;][]{2019MNRAS.483..744T, 2019MNRAS.487.5416T, 2019MNRAS.490.3196P}.
The advent of the James Webb Space Telescope (JWST) has recently enabled tests in the high-redshift Universe, with its unprecedented resolution and sensitivity at infrared wavelengths desirable for probing stellar structures.
However, several latest works using JWST data show seemingly the opposite to the simulations: visually identified disc galaxies (of typical flattened shapes and nearly exponential surface brightness profiles) dominate the galaxy populations beyond cosmic noon \citep[e.g.,][]{2023ApJ...946L..15K, 2023ApJ...955...94F, 2024ApJ...960..104S}.
These empirical results have been extensively interpreted as that rotation supported and dynamically cold galaxies are prevailing even at high redshifts.
If such interpretation is right, galaxy formation theories in high density regimes may need to be revised.

Nevertheless, inferring the intrinsic morphology and kinematics from imaging alone is inherently uncertain.
Particularly, the flattened shapes of galaxies, which have been deemed indicative of discs in visual morphological classifications, do not necessarily guarantee rotating oblate disc structures.
In addition to ordered rotation, flattening can also be caused by the anisotropy of random motion.
This is the case among nearby giant elliptical galaxies
which were originally thought to be flattened by rotations but later found to be by velocity anisotropy \citep{1975ApJ...200..439B, 1976MNRAS.177...19B, 1977ApJ...218L..43I, 1978MNRAS.183..501B}.
When projected on the sky, elongated galaxies of intrinsic prolate shapes, dynamically supported by radial motions, tend to show themselves in flattening seemingly like the edge-on discs \citep[e.g.,][]{2015MNRAS.453..408C}.

Without deep integral field spectroscopy, it cannot be pinned down whether the prevailing high-$z$ discs are actually supported by anisotropic random motion.
But before that consuming data, in this work we study the stellar kinematics of local galaxies with masses comparable to that of the typical high-$z$ galaxies observed by JWST ($\mathrm{lg}(\mathcal{M}_{\star}/\mathcal{M}_{\odot})\sim9$).
This sets a benchmark of the rotational support of low-mass systems in relatively quiescent formation, providing key information for assessing the case at high $z$ when galaxy assembly is substantially more violent.
In the last section, we bring together other pieces of evidence from observations and simulations to discuss the nature of these high-$z$ flattened galaxies.

Throughout this work we adopt the concordance cosmology with $H_0$ = $70\ \mathrm{km}\,\mathrm{s}^{-1}\,\mathrm{Mpc}^{-1}$, $\Omega_\mathrm{m}$ = 0.3, and $\Omega_{\Lambda}$ = 0.7.

\section{Sample}\label{sec:samp}

The local galaxy samples to be compared with high-$z$ galaxies are drawn from the complete MaNGA survey \citep{2015ApJ...798....7B, 2022ApJS..259...35A}, which ensures better understanding of galaxy shapes with integral field spectroscopy.
MaNGA has observed $\sim10000$ galaxies, the largest survey of the kind, in the local Universe ($0.01<z<0.15$) at angular resolution 2.5 arcsec in FWHM with the integral field units of effective diameters ranging from 12 to 32 arcsec \citep{2015AJ....149...77D}.
This spatial coverage in most of cases ensures observations out to at least 1.5 effective radii of the galaxies.
The spectra cover 360-1030 nm with median instrument broadening $\sigma_{\mathrm{inst}}\,\sim\,72\,\mathrm{km}\,\mathrm{s}^{-1}$ \citep{2016AJ....152...83L} and typical spectral resolution $R \sim 2000$.

We use the spin parameter $\lambda_{R_{\rm e}}$ \citep{2007MNRAS.379..401E}, the effective specific angular momentum of stars, to quantify the rotational support of galaxies and to constrain their intrinsic shapes.
The robust $\lambda_{R_{\rm e}}$ measurements for the complete MaNGA sample are described in \citet{2023ApJ...950L..22W}.
We apply careful quality control to remove systems with spectroscopic or photometric problems, and to exclude merging galaxies in chaotic morphology for which quantification of internal kinematics is not attainable (see more details in \citet{2023ApJ...950L..22W}).
The smearing of rotation field by seeing is corrected for using the analytic functions derived in \citet{2018MNRAS.477.4711G}, which has been tested in simulations \citep{2019MNRAS.483..249H} showing the correction is effective with little systematics.

The flattening of the apparent galaxy shapes is quantified via ellipticity $\varepsilon$, i.e. $1-b/a$, inside the half-light isophote.
Ellipticity $\varepsilon$ together with the position angle of photometric major axis $\mathrm{PA}_\mathrm{phot}$ are determined via Multi-Gaussian Expansion\footnote{Using the \textsc{MgeFit} Python software package of \citet{2002MNRAS.333..400C} available at https://pypi.org/project/mgefit/.} \citep{1994A&A...285..723E} modelling of the SDSS $r$-band surface brightness maps.
The position angle of kinematic major axis $\mathrm{PA}_\mathrm{kine}$ is defined by the axis aligned with the main sense of rotation in the stellar mean velocity field, which is measured by using \textsc{PaFit}\footnote{Available at https://pypi.org/project/pafit/.} \citep{2006MNRAS.366..787K}.
We take the values of ellipticity and the two position angles from the catalogue\footnote{https://manga-dynpop.github.io} of \citet{2023MNRAS.522.6326Z}.
The S\'ersic indices measured for MaNGA galaxies using SDSS optical images are taken from the PyMorph photometric catalogue\footnote{https://www.sdss4.org/dr17/data\_access/value-added-catalogs/?vac\_id=manga-pymorph-dr17-photometric-catalog} \citep{2019MNRAS.483.2057F, 2022MNRAS.509.4024D}.

We retrieve total stellar mass and star formation rate (SFR) for MaNGA galaxies from the version X2 of GALEX-SDSS-WISE Legacy Catalogue\footnote{https://salims.pages.iu.edu/gswlc/} \citep[GSWLC-X2,][]{2016ApJS..227....2S,2018ApJ...859...11S}, which are derived by modelling the ultraviolet, optical, and mid-infrared fluxes.

Among the representative local galaxy populations sampled by MaNGA survey, we cull the following two subsamples for our purpose of suggesting the probable disconnection between exponential flattened shapes and cold rotating discs in JWST high-$z$ galaxies.

\textbf{Low-mass galaxies at $z=0$:}
These are 149 MaNGA galaxies in a narrow stellar mass bin $8.9<\mathrm{lg}(\mathcal{M}_{\star}/\mathcal{M}_{\odot})<9.1$ with robust measurements of stellar spin $\lambda_{R_{\rm e}}$.
Their low masses are comparable to the median masses of JWST high-$z$ samples studied in \citet{2023ApJ...946L..15K, 2023ApJ...955...94F}, thus making them local analogs (in terms of stellar mass) of typical JWST galaxies found in the early Universe.
No environment restriction is applied.
Because these low-mass galaxies are in the low luminosity regime of the MaNGA sample, they are predominantly actively star-forming (the same also for JWST galaxies).
The images and maps of mean velocity and velocity dispersion have been visually checked to make sure the photometric and kinematic measurements are sensible.
We note that the uncertainty in estimating the velocity dispersion below the instrument resolution has been largely reduced \citep{2017MNRAS.466..798C}.
For some pixels, the intrinsic velocity dispersion of stars is negligible, and the broadening of spectral features is predominantly due to the limited velocity resolution of instrument, leading to estimated quadratic velocity dispersion unphysically negative.
The velocity dispersion values in such pixels have been set to zero.
Despite the uncertainty in measuring these intrinsically low values of velocity dispersion, such pixels are quite minor and unimportant in the low-mass sample because at median value they only comprise $\sim4\%$ of all pixels within the area for calculating $\lambda_{R_{\rm e}}$.

\textbf{Milky Way analogs at $z=0$:}
We also select Milky Way analogs in MaNGA survey to represent the real dynamically cold discs in the local Universe.
These 221 analogs are defined by stellar mass range $10.6<\mathrm{lg}(\mathcal{M}_{\star}/\mathcal{M}_{\odot})<10.8$ (around the median of the estimated range in \citet{2023arXiv230316217P}) and are confined within the span of star formation main sequence (SFMS) of local galaxies \citep[section 2.1 of ][]{2023ApJ...950L..22W}.
We do not additionally apply any morphology cut as these galaxies are already overwhelmed by fast-rotating discs with high rotational support \citep{2023NAL000W}.

\vspace{0.25cm}

\section{The projection of ellipsoids}\label{sec:Eproj}

To get a first-order assessment of galaxy intrinsic shapes, later in this paper we will see distributions of the ellipticity $\varepsilon$ of galaxy projected shapes as compared to the analytic predictions for projected ellipsoids.

The projection of a general ellipsoid with three principal axes $c\leq b \leq a$ and two axial ratios $\xi \equiv c/a$ and $\zeta \equiv b/a$ is described in \citet{1985MNRAS.212..767B}, and for convenience we summarize the relevant part below.

The projected axial ratio $q$ (i.e., $1-\varepsilon$) is given by:

\begin{equation}
q(\theta , \phi ; \xi , \zeta)= \left[ \frac{A+C-\sqrt{(A-C)^2+B^2}}{A+C+\sqrt{(A-C)^2+B^2}} \right] ^{1/2}
\end{equation}
where $(\theta, \phi)$ define the line of sight and are the polar angle and azimuthal angle with respect to the shortest and longest axis respectively, and:

% \begin{align}
% &A\equiv \frac{\mathrm{cos}^2\theta}{\xi^2} \left(\mathrm{sin}^2\phi + \frac{\mathrm{cos}^2\phi}{\zeta^2}\right) + \frac{\mathrm{sin}^2\theta}{\zeta^2}\ , \\[6pt]
% &B\equiv\mathrm{cos}\theta \, \mathrm{sin}2\phi \left(1-\frac{1}{\zeta^2}\right)\frac{1}{\xi^2}\ , \\[6pt]
% &C\equiv \left(\frac{\mathrm{sin}^2\phi}{\zeta^2}+\mathrm{cos}^2\phi\right)\frac{1}{\xi^2}\ .
% \end{align}

\begin{equation}
A\equiv \frac{\mathrm{cos}^2\theta}{\xi^2} \left(\mathrm{sin}^2\phi + \frac{\mathrm{cos}^2\phi}{\zeta^2}\right) + \frac{\mathrm{sin}^2\theta}{\zeta^2}\ ,
\end{equation}

\begin{equation}
B\equiv\mathrm{cos}\theta \, \mathrm{sin}2\phi \left(1-\frac{1}{\zeta^2}\right)\frac{1}{\xi^2}\ ,\ \ \ \ \ \ \ \ \ \ \ \ \
\end{equation}

\begin{equation}
C\equiv \left(\frac{\mathrm{sin}^2\phi}{\zeta^2}+\mathrm{cos}^2\phi\right)\frac{1}{\xi^2}\ .\ \ \ \ \ \ \ \ \ \ \ \ \ \ \ \ \ \
\end{equation}

\vspace{0.25cm}

\section{Results}\label{sec:resu}

\begin{figure*}
  \begin{center}
    \includegraphics[width=0.7\textwidth]{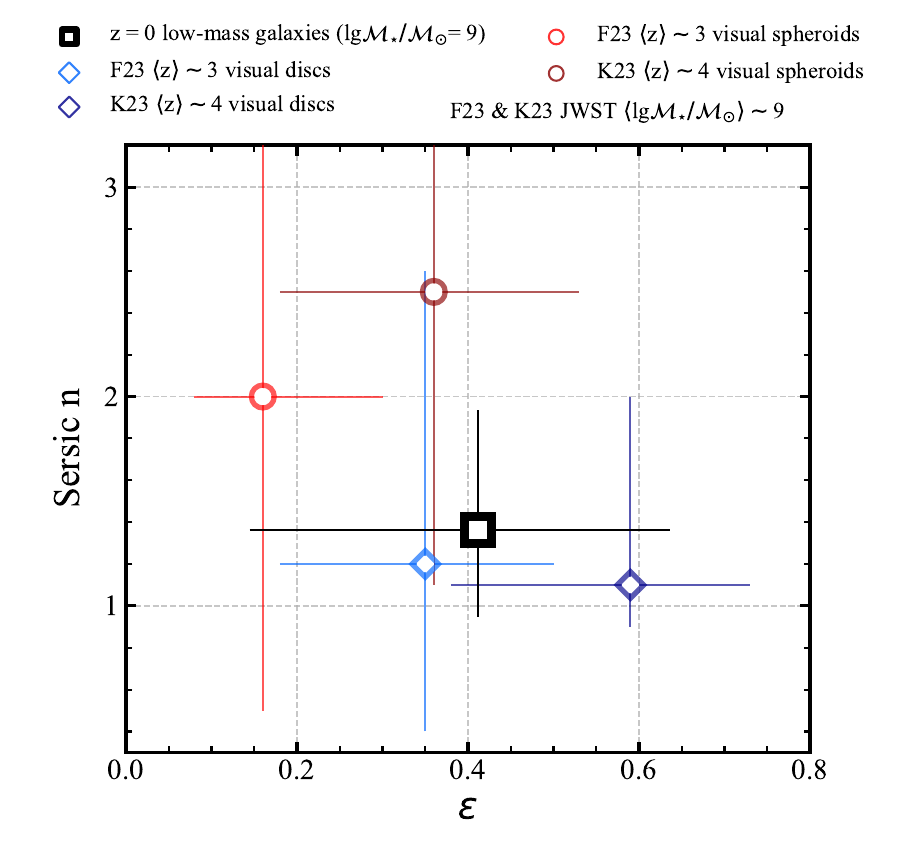}
  \end{center}
 \caption{The concentration of restframe optical surface brightness profiles, as quantified by S\'ersic indices, versus the ellipticity $\varepsilon$ (i.e. flattening or fractional axial difference $1-b/a$) of optical isophotes.
 The red and blue symbols mark the medians and 16th-84th percentiles of JWST visually identified spheroids and discs respectively from \citet{2023ApJ...946L..15K} and \citet{2023ApJ...955...94F}, while the black symbol is for low-mass galaxies in the local Universe observed in MaNGA survey.
 The stellar masses of these low-mass galaxies at $z=0$ are typical in the JWST samples.
 }
 \label{fig:1}
\end{figure*}

JWST CEERS survey \citep{2023ApJ...946L..13F} has imaged thousands of high-$z$ galaxies in the rest-frame optical in the redshift range $2<z<5$ mainly with stellar masses $8<\mathrm{lg}(\mathcal{M}_{\star}/\mathcal{M}_{\odot})<10$.
Statistical analyses of stellar structure show high fractions of galaxies with flattened shapes and low concentrations (nearly exponential profiles), which tend to be visually classified as ``discs''.

In \autoref{fig:1}, we illustrate the galaxy distributions on the S\'ersic n vs. ellipticity parameter plane.
The red and blue symbols mark the medians and 16th-84th percentiles of JWST visually identified spheroids and discs respectively, extracted from two representative works \citep[][abbreviated as K23 and F23 hereafter]{2023ApJ...946L..15K, 2023ApJ...955...94F}.
The sample studied by K23 are restricted to higher redshifts ($z>3$) and larger stellar masses ($9<\mathrm{lg}(\mathcal{M}_{\star}/\mathcal{M}_{\odot})$).
The morphological classifications are also more detailed in K23, and we pick their ``disc only'' objects to represent the population that seem to be dominated by disc structure.
There are systematic offsets between K23 and F23 results, while in general visually classified discs have lower S\'ersic indices and more flattened shapes.
The black symbol shows the distributions for low-mass galaxies in the local Universe, which overlap mostly with the two JWST visual disc populations.
If these low-mass galaxies at $z=0$ were analyzed using images at spatial resolution (e.g., FWHM of point spread function normalized by galaxy size) comparable to that of JWST galaxies, many of them may also tend to be visually classified as ``discs''.

\begin{figure*}
 \includegraphics[width=0.95\textwidth]{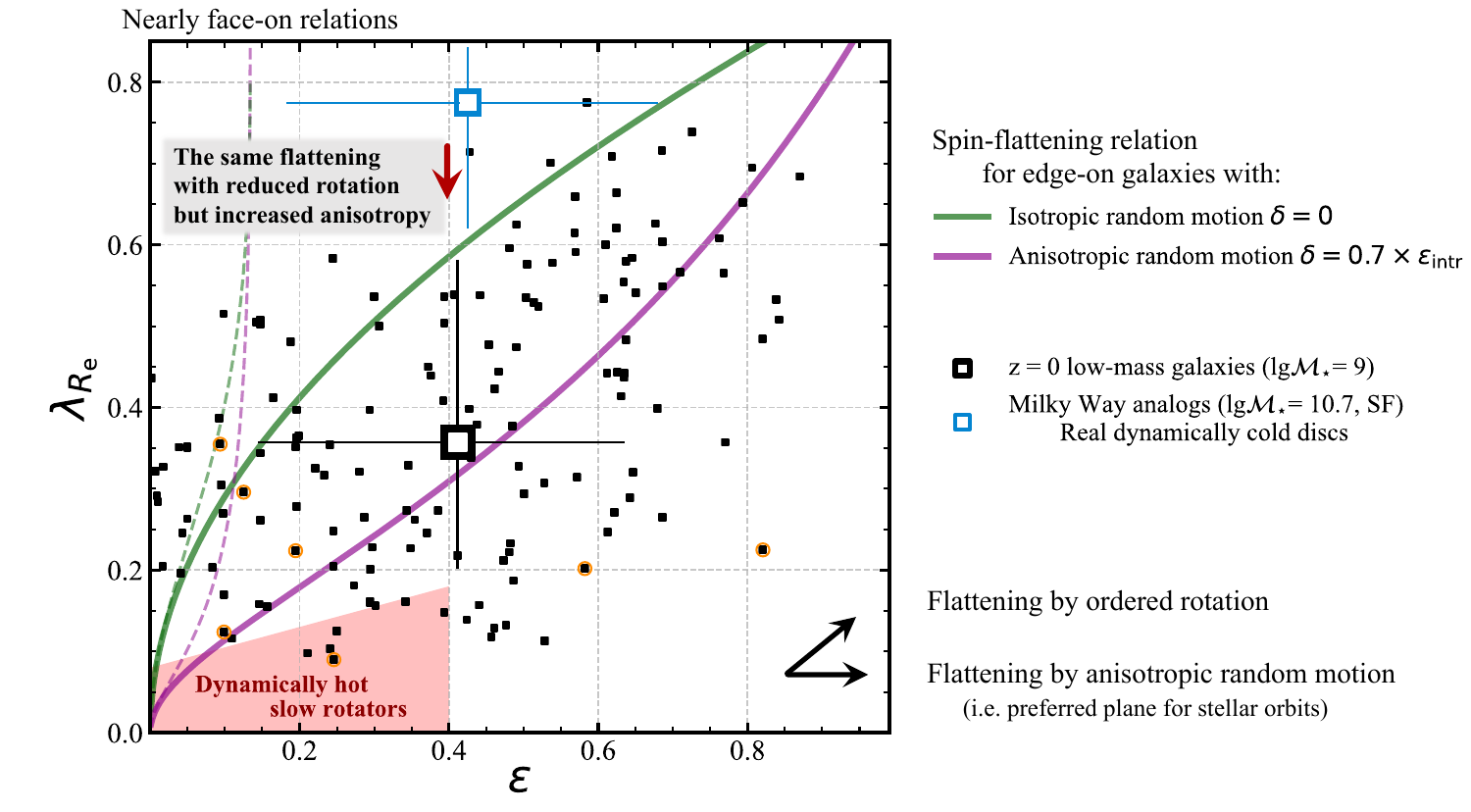}
 \caption{The stellar spin $\lambda_{R_{\rm e}}$ versus flattening $\varepsilon$, indicating the overall warm/hot stellar kinematics of low-mass galaxies in the local Universe.
 The large black square and error bars denote the median and 16th-84th percentiles of the low-mass sample, each galaxy of which is shown by a small black square.
 The most actively star-forming ones (0.5 dex above the SFMS) are circled in orange, and among local low-mass galaxies they can be better analogs of high-$z$ low-mass systems for further similarity in star formation and gas properties.
 For comparison, Milky Way analogs in the local Universe, as representatives of real dynamically cold thin discs, are shown by the large blue square, and the area occupied by dynamically hot slow rotators is illustrated by the shaded trapezoid.
 The green and magenta solid curves are loci respectively for isotropic and anisotropic edge-on axisymmetric galaxies, predicted by tensor virial theorem, with anisotropy in the stellar random motion as given by the relations $\delta=0$ and $\delta=0.7\varepsilon_\mathrm{intr}$.
 The relations for nearly face-on (at an inclination angle of 30 degrees) rotators are shown by the green and magenta dashed curves.
 We note that roughly a half of low-mass galaxies distribute below the magenta solid curve, suggesting that they are significantly more anisotropic than galaxies of higher stellar masses which lie mostly above the magenta solid curve.
 The suggested high anisotropy of low-mass galaxies explains their flattened shapes at low levels of rotation.
 }
 \label{fig:2}
\end{figure*}

However, the measurements of stellar spin indicate that these low-mass galaxies at $z=0$ are dynamically too hot with too little rotational support to be considered a population of cold rotating discs.
\autoref{fig:2} shows the observed galaxy distributions and theoretical predictions on the $\lambda_{R_{\rm e}}$ vs. $\varepsilon$ (stellar spin vs. flattening) diagram.
Again, the large black square and associated error bars are the median and 16th-84th percentiles of low-mass galaxies at $z=0$, while small black squares stand for individual galaxies.
For low-mass galaxies, the median magnitude of stellar spin is 0.35, meaning that the specific angular momenta of these systems are only about $1/3$ of the cases when they are maximally rotation supported.
In terms of rotational support, they are closer to being dynamically hot slow rotators (e.g., giant elliptical galaxies) that occupy the lower left area \citep{2016ARA&A..54..597C} on the $\lambda_{R_{\rm e}}$ vs. $\varepsilon$ plane, than being the real thin discy galaxies like Milky Way analogs (the blue square).

Despite much less rotation, low-mass galaxies are still in shapes that look similarly flattened as those dynamically cold discs, as indicated by their entirely overlapped percentiles of $\varepsilon$ distributions.
This suggests that low-mass galaxies at $z\sim0$ are flattened, to a large extent, by the anisotropy in random motion.

In \autoref{fig:2}, we show how $\lambda_{R_{\rm e}}$ varies with $\varepsilon$, as predicted\footnote{See eqs.14--15 of \citet{2016ARA&A..54..597C}.
The prediction of $\lambda_{R}$ uses the tight relation between $\lambda_{R}$ and $V/\sigma$ derived from observations and two-integral Jeans models \citep[][eq. B1]{2011MNRAS.414..888E}, which has been checked for the low-mass MaNGA galaxies.} by tensor virial theorem \citep{2005MNRAS.363..937B}, for {\it edge-on} axisymmetric galaxies at different velocity anisotropy:
The green solid curve is for isotropic rotators, i.e. those that have equivalent random motion along different directions, and it shows the steady increase of the flattening and rotational support that are in tandem (i.e. $\nearrow$ trend on the $\lambda_{R_{\rm e}}$-$\varepsilon$ diagram: flattening by ordered rotation).
The magenta solid curve is for extreme anisotropy observed among more massive galaxies with $\mathrm{lg}(\mathcal{M}_{\star}/\mathcal{M}_{\odot})>9.5$, following the anisotropy relation of $\delta=0.7\varepsilon_\mathrm{intr}$ \citep{2007MNRAS.379..418C}, where $\varepsilon_\mathrm{intr}$ is $\varepsilon$ viewed edge-on.
The global anisotropy parameter is defined as $\delta\equiv 1-(\sigma_z/\sigma_x)^2$ with $z$ aligned with the symmetry axis of the axisymmetry of galaxies and $x$ being any direction orthogonal to it.
At given rotation, galaxies can be further flattened by increased anisotropy ($\rightarrow$ trend), or the same flattening can be achieved with reduced rotation but increased anisotropy (\textcolor{red}{$\downarrow$} trend).
Projection lowers both $\lambda_{R_{\rm e}}$ and $\varepsilon$ and the isotropic and anisotropic relations at an inclination angle of 30 degrees are shown separately by the green and magenta dashed curves.

In the local Universe, unlike galaxies of these low masses $\mathrm{lg}(\mathcal{M}_{\star}/\mathcal{M}_{\odot})\sim9$, more massive galaxies with regular rotation field are well confined above the magenta solid curve \citep{2007MNRAS.379..418C, 2018NatAs...2..483V, 2020MNRAS.495.1958W}, suggesting this curve to be an empirical anisotropy upper limit.
Rare exceptions are those that manifest two peaks in the velocity dispersion field, which reveal the presence of two counter-rotating disc components \citep{2022MNRAS.511..139B}.
This empirical anisotropy upper limit (i.e. the magenta solid curve) also applies to the Milky Way analogs here, whose distribution can be described as oblate rotators with anisotropy in the range $[0, 0.7\varepsilon_\mathrm{intr}]$ projected at random inclinations.

\begin{figure*}
  \begin{center}
    \includegraphics[width=0.95\textwidth]{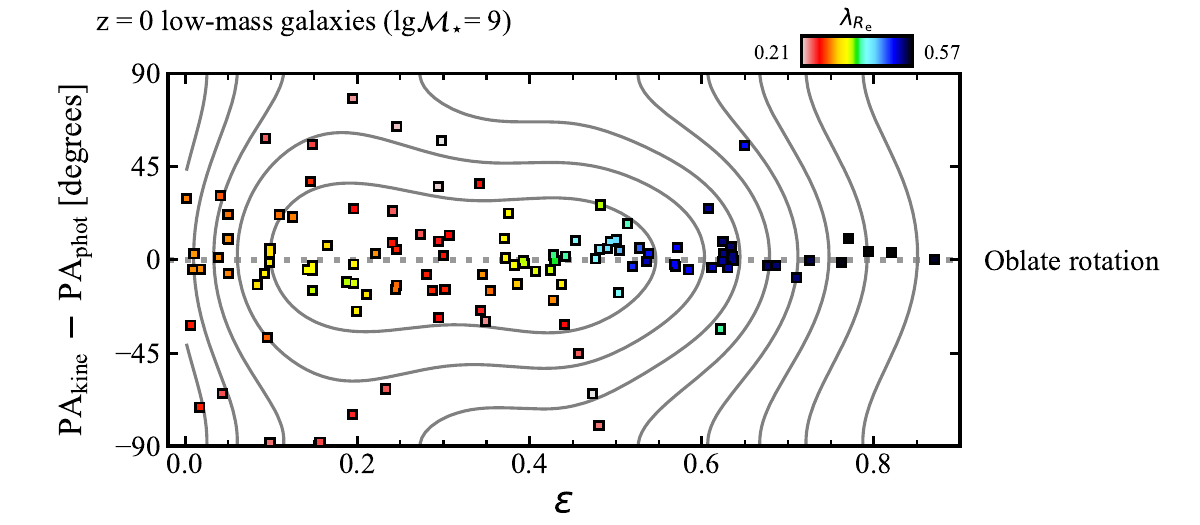}
  \end{center}
 \caption{The misalignment ($\mathrm{PA}_\mathrm{kine}-\mathrm{PA}_\mathrm{phot}$) between kinematic and photometric major axis as a function of ellipticity $\varepsilon$ for MaNGA low-mass galaxies at $z=0$, color coded by stellar spin $\lambda_{R_{\rm e}}$ (slightly smoothed using the locally weighted regression method LOESS by \citet{Cleveland1988} as implemented by \citet{2013MNRAS.432.1862C}; The Python package \textsc{loess} is available at https://pypi.org/project/loess/).
 The measurements of the position angles of the two axes are detailed in the text.
 The gray contour lines show the Gaussian kernel estimated density of galaxies, and to avoid edge effects and account for the periodicity of angles, we have utilized two more duplicates of the data points shifted by $+/-$ 180 degrees respectively.
 Taking measurement uncertainty into account, the distribution is consistent with the case of oblate rotators which have aligned kinematic and photometric major axes.
 }
 \label{fig:3}
\end{figure*}

But at masses as low as $\mathrm{lg}(\mathcal{M}_{\star}/\mathcal{M}_{\odot})\sim9$, there appears to be a large fraction that break this empirical anisotropy upper limit, and lie much below the magenta solid line.
Using the theoretical zero rotation limit\footnote{See the black dashed line in Fig. 3 of \citet{2021MNRAS.500L..27W}.} (i.e. flattening is totally due to anisotropy; theoretically maximal anisotropy $\delta_\mathrm{max}(\varepsilon_\mathrm{intr})$ at given $\varepsilon_\mathrm{intr}$) as reference, the anisotropy relation $\delta(\varepsilon_\mathrm{intr})=0.7\varepsilon_\mathrm{intr}$ assumed by the magenta solid line approximates 80\% of $\delta_\mathrm{max}(\varepsilon_\mathrm{intr})$, which represents the extreme among more massive galaxies.
The distribution lower boundary of low-mass galaxies suggests that their anisotropy extreme approaches 99\% of $\delta_\mathrm{max}(\varepsilon_\mathrm{intr})$, close to the zero rotation limit.
Without the intention of overfitting the data, we tentatively build a mock sample of randomly oriented axisymmetric galaxies with $\varepsilon_\mathrm{intr}$ uniformly sampled in the range $[0.4,0.95]$ and anisotropy $\delta$ uniformly sampled in $[0.7\varepsilon_\mathrm{intr}, 0.99\delta_\mathrm{max}]$.
The distribution of this mock sample on $\lambda_{R_{\rm e}}$-$\varepsilon$ diagram largely resembles the observed distribution of the low-mass galaxies.
Kolmogorov-Smirnov tests\footnote{The Python implementation of the two-dimensional Kolmogorov-Smirnov testing \textsc{ndtest} available at https://github.com/syrte/ndtest.} suggest that it is very likely (p-value: 0.4-0.5) that this mock and real samples are drawn from the same underlying distribution.
The tests rule out the possibility that most of the low-mass galaxies have anisotropy $\delta$ smaller than $0.7\varepsilon_\mathrm{intr}$, which is also clearly indicated by the distribution showing half of them are below the magenta solid line.

Such high anisotropy in the random motion explains the significant flattening of low-mass galaxies at low levels of stellar rotation.
Under the assumption of oblate velocity ellipsoid (i.e. $\sigma_{\phi}\approx\sigma_R$), which is reasonable for more massive regular-rotator galaxies with $\mathrm{lg}(\mathcal{M}_{\star}/\mathcal{M}_{\odot})>9.5$ \citep{2007MNRAS.379..418C, 2016ARA&A..54..597C}, \citet{2021MNRAS.500L..27W} finds that, beyond this empirical anisotropy upper limit, the solutions to axisymmetric Jeans equations are unphysical in realistic galaxy potentials.
This suggests that real galaxies can hardly establish hydrostatic equilibrium at that large anisotropy without biased stellar motion on the equatorial plane.
Here, the fact that this empirical upper limit does not hold for low-mass galaxies implies that these low-mass galaxies, many of which show irregularity in velocity field, tend to possess the non-oblate velocity ellipsoid ($\sigma_{\phi} \not\approx \sigma_R$) observed among galaxies with nonregular rotation \citep[Fig. 11 of][]{2016ARA&A..54..597C}.

Different types of orbital anisotropy can lead to different intrinsic shapes.
\autoref{fig:3} shows that the photometric and kinematic major axes\footnote{Both position angles have been visually checked to exclude clearly unreliable measurements, such as those due to large area of low signal-to-noise ratio in the velocity maps.} of local low-mass galaxies are statistically aligned with each other, with expected large scatter around zero $\mathrm{PA}_\mathrm{kine}-\mathrm{PA}_\mathrm{phot}$ at low $\varepsilon$ when the photometric major axis is not well defined.
If galaxies are intrinsically prolate or triaxial, the photometric and kinematic major axes are not usually aligned, and the general alignment observed among the local low-mass galaxies indicates their overall oblate shapes in axisymmetry (see a more detailed review in Section 3.3 of \citet{2016ARA&A..54..597C}).

The oblate shapes and large velocity anisotropy of these local low-mass systems mean that they have a substantial amount of kinetic energy of random motion along the equatorial plane, in either the radial or tangential direction.
Mergers are commonly found reasons for radial anisotropy on galactic scales, such as for the Gaia Enceladus \citep{2018MNRAS.478..611B} in Milky Way, but are of limited importance in low-mass galaxies due to their low \textit{ex situ} mass fractions \citep[e.g.,][]{2019MNRAS.487.5416T}.
Moreover, fine tuned merger configuration might be required to confine the accreted stars in the equatorial plane.
Tangentially biased orbits ($\sigma_{\phi}>\sigma_R$) on the other hand, like the special cases of significant counterrotating discs seen among more massive galaxies (Fig. D1 of \citet{2020MNRAS.495.1958W} in the appendix), are seemingly more realistic and practical for the shapes and anisotropy of these local low-mass galaxies.
Due to frequent angular momentum changes of the feeding cosmic-web streams, the spin axes of gas discs in low-mass galaxies flip in less than an orbital time \citep{2020MNRAS.493.4126D}.
Therefore, low-mass galaxies may indeed host significant counterrotating stellar populations.
But we note that their small intrinsic velocity dispersion well below the MaNGA instrumental dispersion, as well as the spatially overlapping counterrotation after multiple spin flips, make the counterrotating signals hard to reveal from the velocity and dispersion maps.

With their particular properties, the local low-mass galaxies form another significant population below the magenta line in \autoref{fig:2}, in addition to the widely acknowledged massive slow rotator population that are made by a series of major and minor dry mergers.
In the following section, we will bring more discussion on these two populations below the magenta line.

\section{Discussion and conclusion}\label{sec:dac}

In this work, we learn from the stellar kinematics and shapes of low-mass, $\mathrm{lg}(\mathcal{M}_{\star}/\mathcal{M}_{\odot})\sim9$, galaxies in the local Universe.
These low-mass systems, mostly star-forming, have nearly exponential profiles of surface brightness and apparent shapes that are comparably flattened as dynamically cold discy galaxies like our Milky Way.
Such photometric features are closely related to the presence of dominant disc structures in more massive populations at $z=0$.
This link, however, is not guaranteed unconditionally.
We show that these low-mass galaxies, which may be visually classified as discs according to their photometric features, are actually much more dominated by stellar random motion than those real rotation-supported discs, and their apparent flattening are attributed to high anisotropy in the random motion.
The anisotropy is most likely to be tangential, i.e. stellar orbits are tangentially biased, so as to account for the intrinsic oblate shapes of these local low-mass galaxies indicated by the general alignment between their photometric and kinematic major axes.

\begin{figure*}
  \begin{center}
    \includegraphics[width=0.85\textwidth]{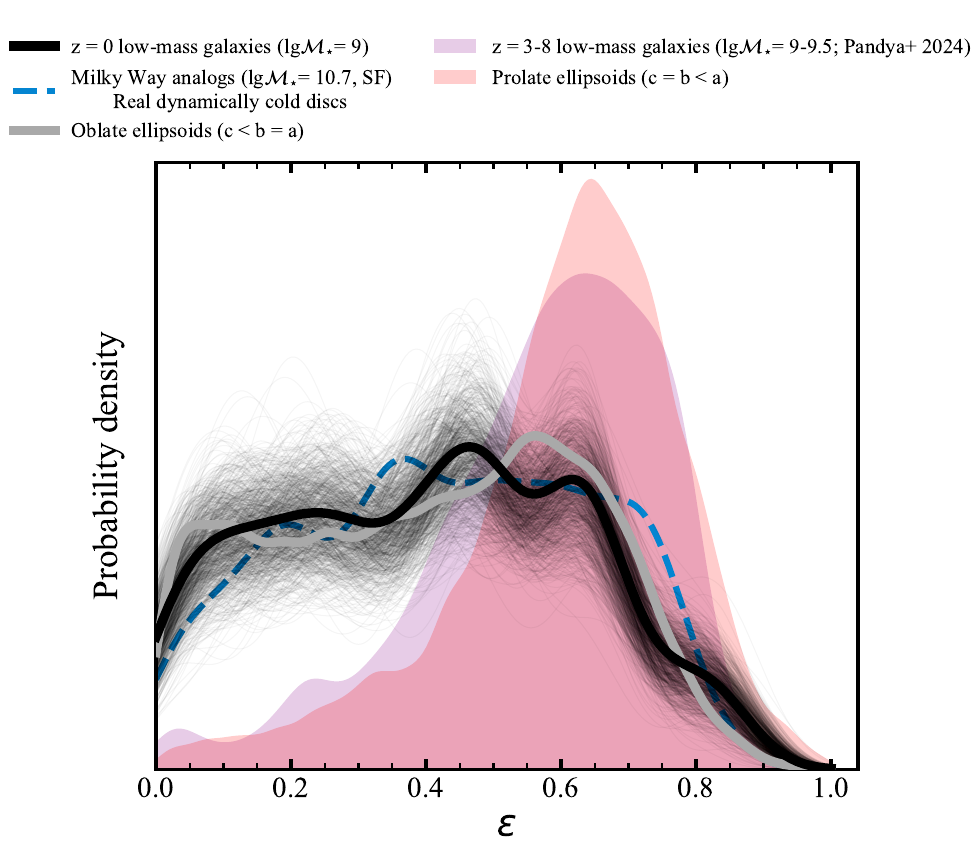}
  \end{center}
 \caption{The probability density distributions of the ellipticity $\varepsilon$ of galaxy projected shapes, for low-mass galaxies (black solid line) and Milky Way analogs (blue dashed line) observed at $z=0$ and for galaxies in the early Universe $z=3-8$ ($9<\mathrm{lg}(\mathcal{M}_{\star}/\mathcal{M}_{\odot})<9.5$ in JWST-CEERS studied by \citet{2024ApJ...963...54P}; the purple shaded area).
 The density distributions are derived via kernel density estimation with Gaussian kernels of widths determined based on Scott's rule.
 For local low-mass galaxies, we estimate the uncertainty of the distribution using 1000 bootstrap samples shown by the thin black lines.
 Suggesting the intrinsic shapes to first order, we have also shown the predicted distributions of oblate ($c<b=a, c/a \sim \mathcal{N}(\mu=0.3,\sigma=0.1)$; the thick gray line) and prolate ($c=b<a, c/a \sim \mathcal{N}(\mu=0.3,\sigma=0.1)$; the orange shaded area) ellipsoids randomly projected on the sky plane.
 }
 \label{fig:4}
\end{figure*}

Below the empirical anisotropy relation (i.e. the magenta line in \autoref{fig:2}) for the axisymmetric oblate regular rotators dominating in the local Universe, the primary population are slow-rotator early-type galaxies at the very high mass end $\mathrm{lg}(\mathcal{M}_{\star}/\mathcal{M}_{\odot}) \gtrsim 11$ \citep{2016ARA&A..54..597C, 2018MNRAS.477.4711G, 2020MNRAS.495.1958W, 2021MNRAS.508.2307V}.
The formation of low-mass galaxies with high tangential anisotropy gives rise to another prominent population below the relation at the low mass end $\mathrm{lg}(\mathcal{M}_{\star}/\mathcal{M}_{\odot}) \sim 9$.
The massive ones are formed\footnote{Non-merger origin may exist but is rare \citep[e.g.,][]{2022MNRAS.509.4372L}.} by a number of major and minor dry mergers, evidenced by their slightly triaxial shapes \citep{2007MNRAS.379..418C, 2011MNRAS.414..888E, 2022ApJ...930..153S}, central stellar cores \citep{1996ApJ...464L.119K, 1997AJ....114.1771F, 2020A&A...635A.129K}, extremely low bulk rotation, massive and extended stellar halos, environment \citep[e.g.,][]{2013ApJ...778L...2C, 2020MNRAS.495.1958W, 2021MNRAS.508.2307V, 2023MNRAS.521.2671S}, and the others.
The low-mass ones, in spite of low angular momenta, still retain the oblate shapes that might be consequence of multiple spin flips of the gas discs \citep{2020MNRAS.493.4126D}.
Spectra of higher velocity resolution are needed to resolve the underlying counterrotating stars in these systems of low velocity dispersion.

Studying kinematics of low-mass galaxies in the local Universe serves as an invaluable proxy for understanding their counterparts recently studied photometrically at high redshift.
Beyond cosmic noon $z>2$, JWST images have revealed a dominating population of galaxies with flattened shapes and nearly exponential surface brightness profiles.
Visual morphological studies classify these galaxies as disc population, and it is widely believed that many of them are dynamically cold discs supported by bulk rotation.
However, we note that the high-$z$ JWST galaxies are mainly star-forming and have typical stellar masses (i.e. the median) of $\mathrm{lg}(\mathcal{M}_{\star}/\mathcal{M}_{\odot})\sim9$, thus making them distant counterparts of the local population presented in this work.
While we show that the local low-mass population are also significantly flattened and have nearly exponential profiles, they are no where near being rotation-supported systems.

From theoretical point of view, the stellar kinematics of actively star-forming galaxies at higher redshifts are governed more by random motion than their local analogs of equal masses \citep[EAGLE, TNG100, TNG50;][]{2019MNRAS.483..744T, 2019MNRAS.487.5416T, 2019MNRAS.490.3196P}, because the fast accumulated gas becomes gravitationally unstable toward star formation far ahead of complete gas cooling and settling down into a disc \citep{2024MNRAS.532.3808M}.
This is consistent with observations that molecular and ionized gas, which stay closely with young stars, are dynamically hotter at higher redshifts \citep[e.g.,][]{2015ApJ...799..209W}.
Significant rotation of cool and cold gas can still be established at extremely high redshifts \citep[e.g.,][]{2024arXiv240218543F, 2024arXiv240506025R}, particularly for massive halos as simulation suggested, but tends to be fragile and destructed after several orbital time \citep{2020MNRAS.493.4126D}.

The typical JWST galaxies observed at high $z$ are thus expected to be supported even more by random motion than the local equal-mass counterparts highlighted in this work which were formed recently and quiescently.
For the local counterparts, if we focus on those with highest SFRs (0.5 dex above the SFMS; orange circles in \autoref{fig:2}), that are more analogous to the high-$z$ population in terms of star formation and gas properties, the median stellar spin does decrease further by 40\%.

There are indeed other pieces of evidence for high-$z$ star-forming galaxies being dynamically warm/hot systems with large velocity anisotropy.
Most notably, high-$z$ galaxies of relatively low masses feature an asymmetric distribution of projected axis ratio that is biased toward small values.
This is the characteristic of intrinsically elongated systems.
\citet{2019MNRAS.484.5170Z} infer the intrinsic shapes of galaxies by modelling the projected axis ratios and sizes based on HST images, and find that in the redshift range $2<z<2.5$, 70\% of $9<\mathrm{lg}(\mathcal{M}_{\star}/\mathcal{M}_{\odot})<9.5$ galaxies and 60\% of $9.5<\mathrm{lg}(\mathcal{M}_{\star}/\mathcal{M}_{\odot})<10$ galaxies are prolate, and the discy population comprise fewer than 20\%.
The prevalence of low-mass prolate galaxies at high redshifts is confirmed by the latest JWST results \citep{2024ApJ...963...54P}, showing that about 70\% of $9<\mathrm{lg}(\mathcal{M}_{\star}/\mathcal{M}_{\odot})<10$ galaxies are prolate at $2<z<8$.
\citet{2024ApJ...963...54P} also find that high-probability prolate and oblate candidates have remarkably similar S\'ersic indices around one and non-parametric morphological properties.
With integral field spectroscopy, our work suggests that these prolate systems, unlike rotation-supported discs in oblate shapes, are governed by anisotropic random motion.

The prolate shapes of high$-z$ low-mass galaxies can naturally be rooted in the shapes and kinematics of their dark matter halos \citep{2019MNRAS.485..972T}.
Simulated low-mass galaxies at high $z$ generally possess elongated prolate shapes, due to the gravitational impact from a dominant dark matter filament that feeds the central galaxy \citep{2015MNRAS.453..408C, 2016MNRAS.458.4477T}, whose torque efficiently drives stars to anisotropic hot orbits.
Recently, \citet{2024ApJ...961...51V} show that, with self-supervised learning algorithm, machine identifies a significant fraction of JWST visually classified disc galaxies as systems similar to the mock TNG50 galaxies with prolate shapes and low angular momenta.
High-$z$ galaxies, including these elongated systems, often show clumpy morphologies of luminous subcomponents that suggest early galaxy formation via frequent mergers \citep[e.g.,][]{2024arXiv240521054A, 2024arXiv240618352H}.
The clumpiness of some of the elongated galaxies thus might indicate their temporary states far from equilibrium, and that their prolate shapes are under fast evolution.

Both having low rotational support, the intrinsic shapes of low-mass galaxies with $\mathrm{lg}(\mathcal{M}_{\star}/\mathcal{M}_{\odot})\sim9$ have transitioned from prolate at $z>2$ to oblate at present.
\autoref{fig:4} illustrates this transition with the ellipticity distributions of local ($z=0$; black line) and distant ($z=3-8$; purple shaded area; taken from \citet{2024ApJ...963...54P}) low-mass galaxies.
Apparently, the projected shapes of local and distant low-mass galaxies are significantly different, with the distant low-mass galaxies showing much more highly flattened shapes on the sky.
The distribution of local low-mass galaxies follows that of the Milky Way analogs (blue dashed line), the real dynamically cold discs, but the distribution of Milky Way analogs is slightly shifted to higher values.
Analytical predictions for randomly oriented oblate (the gray line) and prolate (orange shaded area) ellipsoids, both assuming a Gaussian distribution $\mathcal{N}(\mu=0.3,\sigma=0.1)$ for the axial ratio $\xi \equiv c/a$, match the observations in remarkable details.
The transition of intrinsic shapes is associated with a change in the type of velocity anisotropy.
While the local low-mass galaxies are suggested to possess tangentially biased stellar orbits, the stellar random motions of high-$z$ low-mass galaxies are radially biased, probably along the predominant feeding filament of dark matter \citep{2015MNRAS.453..408C, 2019MNRAS.488.5580P, 2024arXiv240717552P}.

To conclude, the high-$z$ populations recently observed by JWST are not dominated by dynamically cold discs supported by rotation, and their significant flattening are driven by anisotropy in stellar random motion.
The ubiquitous florescence of rotation-supported discs may still remain a low-$z$ phenomenon.
JWST integral field spectroscopy for statistically significant samples of high-$z$ galaxies  will help to pin down their kinematic nature and intrinsic shapes in the future.

\section*{Acknowledgements}

We appreciate the thoughtful and constructive comments by our reviewer.
YP and BW acknowledge the National Science Foundation of China (NSFC) Grant No. 12125301, 12192220, 12192222, and the science research grants from the China Manned Space Project with No. CMS-CSST-2021-A07.

Funding for the Sloan Digital Sky Survey IV has been provided by the Alfred P. Sloan Foundation, the U.S. Department of Energy Office of Science, and the Participating Institutions. SDSS acknowledges support and resources from the Center for High- Performance Computing at the University of Utah. The SDSS website is \url{www.sdss.org}.

SDSS-IV is managed by the Astrophysical Research Consortium for the
Participating Institutions of the SDSS Collaboration including the
Brazilian Participation Group, the Carnegie Institution for Science,
Carnegie Mellon University, the Chilean Participation Group, the French Participation Group, Harvard-Smithsonian Center for Astrophysics,
Instituto de Astrof\'isica de Canarias, The Johns Hopkins University, Kavli Institute for the Physics and Mathematics of the Universe (IPMU) /
University of Tokyo, the Korean Participation Group, Lawrence Berkeley National Laboratory,
Leibniz Institut f\"ur Astrophysik Potsdam (AIP),
Max-Planck-Institut f\"ur Astronomie (MPIA Heidelberg),
Max-Planck-Institut f\"ur Astrophysik (MPA Garching),
Max-Planck-Institut f\"ur Extraterrestrische Physik (MPE),
National Astronomical Observatories of China, New Mexico State University,
New York University, University of Notre Dame,
Observat\'ario Nacional / MCTI, The Ohio State University,
Pennsylvania State University, Shanghai Astronomical Observatory,
United Kingdom Participation Group,
Universidad Nacional Aut\'onoma de M\'exico, University of Arizona,
University of Colorado Boulder, University of Oxford, University of Portsmouth,
University of Utah, University of Virginia, University of Washington, University of Wisconsin,
Vanderbilt University, and Yale University.

\section*{Data Availability}

The measurements and quality assessments of stellar spin $\lambda_{R_{\rm e}}$ can be found in Table 1 of \citet{2023ApJ...950L..22W}.
The measurements of ellipticity, photometric and kinematic position angles $\mathrm{PA}_\mathrm{phot}$ and $\mathrm{PA}_\mathrm{kine}$ are taken from the catalogue of \citet{2023MNRAS.522.6326Z} available at https://manga-dynpop.github.io.
The S\'ersic indices are taken from the PyMorph photometric catalogue \citep{2019MNRAS.483.2057F, 2022MNRAS.509.4024D}, published on https://www.sdss4.org/dr17/data\_access/value-added-catalogs/?vac\_id=manga-pymorph-dr17-photometric-catalog.
Stellar mass and SFR of MaNGA galaxies are retrieved from GSWLC-X2 catalogue \citep{2016ApJS..227....2S,2018ApJ...859...11S} available on https://salims.pages.iu.edu/gswlc/.

%% For this sample we use BibTeX plus aasjournals.bst to generate the
%% the bibliography. The sample631.bib file was populated from ADS. To
%% get the citations to show in the compiled file do the following:
%%
%% pdflatex sample631.tex
%% bibtext sample631
%% pdflatex sample631.tex
%% pdflatex sample631.tex

% \bibliography{PaperVII}{}
\bibliographystyle{aasjournal}

\end{document}